\documentclass[11pt,a4paper]{article}
\usepackage[utf8]{inputenc}
\usepackage[english]{babel}
\usepackage[T1]{fontenc}
\usepackage{url}            
\usepackage{booktabs}
\usepackage{amsmath}
\usepackage{amsfonts}
\usepackage{nicefrac}       
\usepackage{microtype}      
\usepackage{lipsum}	
\usepackage{amssymb}
\usepackage{pdfpages}
\usepackage{lmodern}
\usepackage{makeidx}
\usepackage{xcolor}
\usepackage{hyperref}
\usepackage{graphicx}
\usepackage[nottoc]{tocbibind}
\usepackage[left=2.5cm,right=3cm,top=2.4cm,bottom=2.5cm]{geometry}
\usepackage{tikz}
\usepackage{multirow}
\title{\textbf{Dependence Modeling and Risk Assessment of a Financial Portfolio with ARMA-APARCH-EVT models based on HACs.}}
\author{\begin{tabular}{cc}
		\textbf{Dodo Natatou Moutari} &	\textbf{Hassane Abba Mallam} \\
		FAST-Université Abdou Moumouni & ~~~~FAST-Université Abdou Moumouni\\
	 (Niger) &(Niger)\\
	
	dnatatumutari@gmail.com	& \\
		& \\
	\end{tabular}\\
	\begin{tabular}{cc}
	\textbf{Diakarya Barro} &	\textbf{Bisso Saley}\\
SEG, Université Thomas Sankara &~~~~FAST-Université Abdou Moumouni\\
(Burkina Faso) &		(Niger)\\
	dbarro2@gmail.com& \\
	&  \\
	\end{tabular}  }
	\date{}

\usepackage{hyperref} 
\hypersetup{pdfstartview=XYZ} %
\begin{document}
	\maketitle
	\begin{abstract}
Multivariate modeling of dependence and its impact on risk assessment remains a major concern for financial institutions. Thus, the copula model, in particular Archimedean hierarchical copulas (HAC) appears as a promising alternative, capable to precisely capture the structure of dependence between financial variables. This study aims to widen the sphere of pratical applicability of the HAC model combined with the ARMA-APARCH volatility forecast model and the extreme values theory (EVT). A sequential process of modeling of the VaR of a portfolio based on the ARMA-APARCH-EVT-HAC model was discussed. The empirical analysis conducted with data from international stock market indices clearly illustrates the performance and accuracy of modeling based on HACs.\\

\end{abstract}
 \section{Introduction}
Quantitative research in finance has focused in recent years on the analysis of the strength of dependence between financial variables. Thus, taking this dependence into account is a canal for better selection of portfolios promoting good risk management. The performance of an asset portfolio is a function of several factors including profitability, the risk of each security and the correlations, that is to say the dependence between these variables. It is therefore inevitable to study the behavior of each security individually in order to assess the contribution of each of the assets making up the portfolio. Thus, the risk assessment of financial portfolios requires a better analysis of the dependence structure.\\

Moreover, according to the financial literature \textcolor{blue}{\cite{B17}}, one of the characteristics of temporary financial series is that the movements of one series can affect those of other series. Thus, several approaches are used to model the structures of dependence between financial variables, one of the most widespread of which is the theory of the copula. \\

The relevance of modeling the dependence between financial variables using copula models has been recognized by the existing literature. The copula model was first proposed in the work of Sklar in 1959. The latter proposed a first theorem which allows the construction of these dependence functions and the decomposition of a multivariate distribution into its marges see for example \textcolor{blue}{\cite{B14}}.\\

Investors have a particular interest in the risk assessment of their portfolio, hence the need to develop tools able to quantify the likely loss and describe the market movements. Several risk measurement tools are available in financial modeling, in particular, the value at risk (VaR). The concept of value at risk (VaR) was first used in the insurance industry. But it was in the 1990s that this measure was made popular by the \textcolor{blue}{JP Morgan} bank, thanks to its \textcolor{blue}{RiskMetrics} system see \cite{B8}. Despite its applicability, VaR offers very limited information. Indeed, \textcolor{blue}{Artzner et al \cite{B25}} noted the shortcomings linked to VaR, notably the fact that it does not meet all the conditions for a coherent risk measurement. hence the need to combine VaR with other statistical risk analysis and management tools such as copulas, ARMA-APARCH volatility analysis models, conditional VaR or approach of extreme values.\\

Financial markets are mainly characterized by periods of high and low volatilities. The study of this volatile character requires a rigorous and relevant modeling strategy. The ARMA-GARCH model and its ARMA-APARCH extension offers many potentialities for analyzing and interpreting volatility. Thus, the use of copulas makes the volatility models more efficient by making possible the choice of the marginal model, that is to say the univariate volatility model, the analysis of the structure of dependence and the construction of other non-elliptical distributions.\\

Several works on the modeling of dependence structures and risk analysis have followed one another. For example \textcolor{blue}{Dorothea Diers and al \cite{A2}} used the copula approach to model the dependence structures of non-life insurance risks. \textcolor{blue}{Natalie Neumeyer and al \cite{A3}} examined the dependence structure of two (or more) time series in the presence of a (possibly multivariate) covariate which may include past values of the time series. \textcolor{blue}{Maziar Sahamkhadam and al \cite{B17}} used simulated returns from the copula model not only in Min-CVaR (minimum conditional value at risk) portfolios, but also for CET (certainty equivalence tangency criteria) and GMV (global minimum variance) portfolios and modeled the dependence structure between stock market returns using elliptical and Archimedean copulas. \textcolor{blue}{Ostap Okhrin and Anastasija Tetereva \cite{B9}} proposed an extension of the work of \textcolor{blue}{Fengler and Okhrin} by introducing the Realized Hierarchical Archimedean Copula (rHAC), which allows more flexibility and is applicable to the management of dimensional portfolios and based estimation of the structure and parameters of the rHAC only on a covariance matrix produced, but the dependence between the financial variables is generally not linear. \textcolor{blue}{Yang Baia and al \cite{B8}} presented a hybrid method for forecasting the overnight VaR for large-dimension portfolios, they used the implementations of multivariate GARCH models and copula methods for illustrate the performance of a VaR prediction modeling process. This document differs from previous similar studies in several aspects. First, we use the simulated returns of the ARMA-APARCH model combined with the EVT-HAC model from parameters estimated with real data for forecasting the overnight VaR for large portfolios. So, we closely examine the ARMA-APARCH-EVT model based on hierarchical Archimedean copulas (HACs) and carry out a comparative study with the model based on classic Archimedean copulas (ACs). The ARMA-APARCH-EVT model  model based on HACs has interesting advantages. Indeed, it overcomes some considerable drawbacks of classical Archimedean copulas (ACs). ACs according to \textcolor{blue}{Wenjun Zhu et al \cite{B3}}, despite having advantages of simplicity, suffer from the anomaly of being a fully exchangeable structure. The HAC model has been proposed to partially tame this anomaly by nesting two or more ACs with appropriate grouping. Second, this study aims to establish a clear picture of the dependence between international stock indices. To our knowledge, this is the first study which links the American and European stock market indices and the African stock index BRVM and analyzes their interactions with the model of the hierarchical Archimedean copula function. We structure the rest of the article as follows. Section 2 recalls the basic tools of the econometric methodology. Section 3 deals with modeling with the ARMA-GARCH model based on the EVT-HAC. Section 4 presents the empirical study and the results. Section 5 is concerned with the conclusion and discution.
\section{Methodology}
The aim of this section is to present the basic econometric tools which offer more flexibility in forecasting, estimating and taking into account dependence.
\subsection{A survey of ARMA-APARCH forecast model}
Within the framework of this study, the behavior of each series of financial returns is analyzed by the ARMA-APARCH models. The ARMA-APARCH model is one of the most important models because it includes different types of models notably ARMA-ARCH, ARMA-GARCH, ARMA-GJR-GARCH, ARMA-TARCH see \cite{C1} for more details. These models are used to analyze and to forecast volatility. Thus, the average equation for each series of returns is modeled by an ARMA(p, q) process while the variance equation is modeled by an APARCH (m, n) process. The general form ARMA(p, q)-APARCH(m, n) can be given by:
\begin{equation}
r_{i,t}=\alpha_{i,0}+\sum_{j=1}^{p}\phi_{i,j}r_{i,t-j}+\sum_{k=1}^{q}\theta_{i,j}\epsilon_{i,t-j}+\epsilon_{i,t} 
\end{equation}
where
\begin{equation}
	\epsilon_{i,t}= z_{i,t}\sigma_{i,t},~~ with~~ z_{i,t}\approx (i.i.d)
\end{equation}
and
\begin{equation}
\sigma_{i,t}^{\delta}=\omega_{i}+\sum_{j=1}^{m}\alpha_{i,j}\left( |\epsilon_{i,t-j}|-\gamma_{i,j}\epsilon_{i,t-j}\right)^{\delta} +\sum_{k=1}^{n}\beta_{i,k}\sigma_{i,t-k}^{\delta}
\end{equation}
while $ r_{i, t} $ denotes the real returns of the asset $ i = 1, \ldots, d $ at time t, $ z_{i, t} $ the error terms also called innovation terms or residual terms (independent identically distributed). A sufficient condition to guarantee the positivity of the variance is that all the parameters $ \omega_{i}, \alpha_{i, j}$ and $ \beta_{i, k}, $ be positive and $ -1<\gamma_{i,j}<1 $ . The process is stationary in the second order if and only if \[ \sum_{j=1}^{m}\alpha_{i,j}\mathbb{E}\left[ \left( |\epsilon_{i,t-j}|-\gamma_{i,j}\epsilon_{i,t-j}\right)^{\delta}\right]  +\sum_{k=1}^{n}\beta_{i,k}\sigma_{i,t-k}^{\delta} < 1.\]

Note that the APARCH process corresponds to:\\
(a): an ARCH process when $ \delta=2 , \gamma_{i}=0$ and $ \beta_{i}=0 $.\\
(b): a GARCH when $ \delta =2 $ and $ \gamma_{i}=0$.\\
(c): a GJR-GARCH model when $ \delta =2$.\\
(d): a T-GARCH model when $ \delta=1 $ and $ \beta_{j}=0$.
\subsection{Overview of copulas models}
Copula functions are mathematical tools which imply enormous applications in statistics and in finance in particular to analyze the structure of dependence between variables. Besides great flexibility in manipulation, Copula functions allow a more realistic representation of dependence by overcoming the shortcomings of the traditional approach based on the linear correlation coefficient. Copulas have interesting properties that enrich and facilitate multidimensional quantitative research. The copula theory is used to not only decompose a multivariate distribution into univariate marginal distributions but also to join results obtained in the univariate case to the multivariate case.\\

Originally proposed by \textcolor{blue}{Sklar, 1959} see \cite{B14}, copulas are multivariate distribution functions whose marginal distributions are uniform over $ I = [0,1] $. These dependence functions give the possibility of relating the joint density of a random vector and its marginal densities. In essence, the results of the theorem of \textcolor{blue}{Sklar, 1959} are as follows. \\

Let $F$ be a distribution whose marginal distributions $ F_1, F_2, \cdots , F_m $ are continuous. The copula C is unique and is defined forall $u_1,\cdots,u_m \in [0,1]^{m}$ by:\\
\begin{equation}
C(u_1,\cdots,u_m) = F(F_1^{-1}(u_1),\cdots ,F_m^{-1}(u_m)).
\end{equation}
where the quantile functions $ F_1^{- 1}, F_2^{- 1}, \cdots $ and $ F_m^{- 1} $ are generalized inverses (continuous to the left) of the functions $ F_i $. \\

The following result gives us the expression of Kendall's tau, using copulas. Considering X and Y two continuous v.a of copula C, then the Kendall tau of X and Y denoted by $ \tau_{X, Y} $ or $ \tau_C $ is defined by: 
\begin{equation}
\tau_{X,Y} = \tau_C = 4\int\int_{I^{2}} C(u,v)dC(u,v)-1 =1- \int\int_{I^{2}}\frac{\partial }{\partial u}C(u,v)\frac{\partial }{\partial v }C(u,v)dudv
\end{equation}
We can interpret $ \tau_C $ as being the average of v.a C(U, V) hence:
\begin{equation}
\tau_C = 4\mathbb{E}[C(U,V)]-1
\label{tau}
\end{equation}

There are several families of copulas adjusted according to the application context. Indeed, any distribution associated with a vector whose marginals are uniform over [0,1] defines a copula. \\

In particular, the family of Archimedean copulas is a class of copulas with extraordinary advantages including, especially the copulas of Gumbel, Clayton, Frank and Joe. These copulas have interesting properties and can be easily constructed and simulated. \\

An extension of the Archimedean copulas class gives rise to hierarchical Archimedean copulas (HACs) first introduced by \textcolor{blue}{Joe \cite{B14}} in 3 dimensions then by \textcolor{blue}{McNeil \cite{B1}} in the d-dimensional case. HACs are also called nested Archimedean copulas. \textcolor{blue}{McNeil \cite{B1}} and \textcolor{blue}{Hofert \cite{B11}} gave an explicit stochastic representation of nested Archimedean copulas and succeeded in developing a fast sampling algorithm for these copulas similar to the \textcolor{blue}{Marshall-Olkin} algorithm for Archimedean copulas see \textcolor{blue}{\cite{B12}}.
\subsubsection*{Archimedean copulas}
The multivariate Archimedean copula class has enjoyed particular success in financial modeling. These copulas are used as powerful parametric models to determine the dependence structure of random vectors. A copula C is said to be Archimedean, if and only if, it is of the form: 
\[ C(u_1,\cdots,u_n) = \phi^{-1}\left(  \sum_{i=1}^{n}\phi(u_i) \right)  ;~~ (u_1,\cdots,u_n) \in [0, 1]^{n} \]
where $ \phi: [0,1] \rightarrow [0, + \infty [$, called Archimedean generator is a strictly decreasing continuous function such that:
$ \phi (1) = 0 $ and $ \phi (0) = + \infty $ and that
$ \phi^{- 1} $ the pseudo-inverse of $ \phi $ is completely monotonic on $[0,+\infty[$ i.e 
\[\forall ~~t \geq 0~~ et ~~ \forall~~ k \geq 0 ,~~  (-1)^{k}(\phi^{-1})^{(k)}(t) \geq 0 ;\]
where $ (\phi^{- 1})^{(k)} $ is the derivative of order k of $ \phi^{- 1} $. For us, it has seem to be necessary to recall here the main models of archimedean copulas.

\begin{table}[htbp]
	\centering
	\caption{Examples of Archimedean copulas most used in the financial market in dimension 2.}
	\begin{tabular}{|c|c|c|}
		\hline 
		Family& Parametric generator $\varphi_{\theta}(t)$ & copula, $C(u_1,u_2)$; $ (u_1,u_2) \in I^{2} $ \\ 
		\hline
		Gumbel & $(-\ln t)^{\theta}$; $\theta \in [1, +\infty[ $& $\exp (-\left[ (-\ln u_1)^{\theta}+(-\ln u_2)^{\theta}\right] ^{1/\theta}$ \\ 
		
		Franck & $-\ln \dfrac{e^{-\theta t}-1}{e^{-\theta}-1}$;  $\theta\neq0$ & $\frac{-1}{\theta }\ln \left[ 1+\dfrac{(e^{-\theta u_1}-1)(e^{-\theta u_2}-1)}{e^{-\theta}-1}\right] $ \\ 
		
		Clayton & $\frac{1}{\theta}(t^{-\theta}-1)$; $ \theta \in [-1, 0[\cup]0, +\infty[$ & $\max \left[ (u_1^{-\theta}+u_2^{-\theta}-1)^{-1/\theta}, 0\right] $\\ 
		Joe & $-\ln\left[ 1-(1-t)^{\theta} \right]$; $\theta \in [1, +\infty[$& $ 1-\left[ (1-u_{1})^{\theta}+(1-u_{2})^{\theta}-(1-u_{1})^{\theta}(1-u_{2})^{\theta} \right]^{1/\theta} $\\
		\hline 
		
	\end{tabular} 
\end{table}
\begin{figure}[htbp]
	\centering
	\includegraphics[scale=0.35]{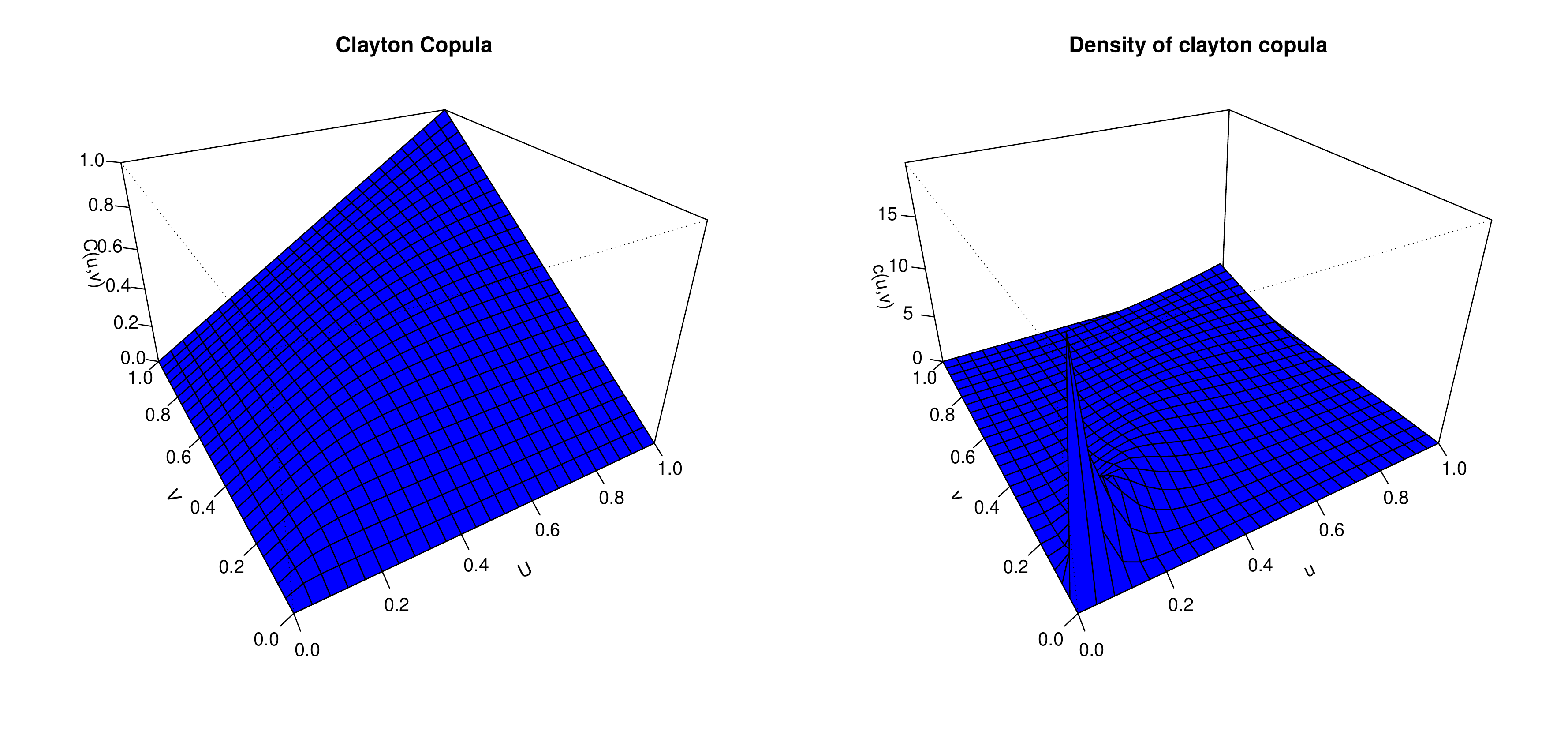}
	\caption{Example of the representation of Clayton's copula and its density with $ \theta = 3 $}
	\label{clay}
\end{figure}
\vspace*{0.1cm}
The Archimedean copula can also be used adequately. Indeed, for this class of copulas, Kendall's tau can be calculated directly from the generator $ \phi $. Considering C the Archimedean copula of two variables X and Y generated by $ \phi $, the Kendall tau of X and Y is given by:
	\begin{equation}\label{tau}
	\tau = 1+4 \int_0^{1}\frac{\phi(t)}{\phi'(t)}dt.
	\end{equation}
	\subsubsection*{Hierarchical Archimedean Copula HAC}
	The basic idea was to build new copulas from Archimedean copulas, in order to widen the number of parameters or to take into account the asymmetry. Applications of nested Archimedean copulas in modeling were introduced by \textcolor{blue}{Joe \cite{B14}, McNeil \cite{B1}}.\\
	
	The class of hierarchical Archimedean copulas (HACs) is of paramount importance insofar as it generalizes the popular class of Archimedean copulas by giving them a hierarchical form. Thus, the dependence structure described by the Archimedean copula is very restricted because it generally depends on a single parameter of the Archimedean generator $ \phi $ and the dependence is symmetrical with respect to the permutation of the variables, that is to say exchangeable. HAC partially overcomes these shortcomings and assigns all random variables in a hierarchical structure with different levels or subgroups. \\
	
	The particular case of the fully nested copula can be given by:
	\begin{equation} \label{HAC}
	C(u_{1}, \ldots,u_{d})
	=C_{1}\{ C_{2}\left( C_{3}\left(u_{1}, \ldots,u_{d-2}\right) u_{d-1}\right) ,u_{d}  \}. 
	\end{equation}
	While using the archimedean generator, it comes that
	\begin{equation}
	C(u_{1}, \ldots,u_{d})  
	= \phi_{1} \{ \phi^{-1}_{1} \circ C_{2}\left( u_{1}, \ldots,u_{d-1}\right)  + \phi^{-1}_{1}  \}
	\end{equation}
	which gives
	\begin{equation}
	 C(u_{1}, \ldots,u_{d})= \phi_{1} \left\lbrace \phi_{1}^{-1}\circ \phi_{2}\left( \phi_{2}^{-1}\left( C_{3}(u_{1},\ldots,u_{d-2})\right) +\phi_{2}^{-1}(u_{d-1})\right) +\phi_{1}^{-1}(u_{d})\right\rbrace.
	\end{equation}
	The composition can be applied recursively using different segmentations of variables leading to more complex HACs. For convenience, let us denote the expression of the structure of a HAC by:  $ \left\lbrace (\ldots(i_{1}\ldots i_{j_{1}}) \ldots) \right\rbrace  $ where $ i_{l} \in  \{1,\ldots,d\} $ is a reordering of the indices of the variables. $ s_ {j} $ denotes the structure of sub-copulas with $ s_ {d} = s $. Further, the d-dimensional hierarchical Archimedean copula is denoted $ C(u_{1},\ldots,u_{d};s, \Theta) $, where $ \Theta $ is the set of parameters of the copula. In fact, the fully nested HAC (\ref{HAC}) can be written in the form:
	\begin{equation}
	C(u_{1},\ldots,u_{d},s=s_{d},\Theta)
	=C\left\lbrace u_{1},\ldots, u_{d}; ((s_{d-1})d),(\theta_{1},\ldots,\theta_{d-1})^{T}
	\right\rbrace.
	\end{equation}
	While using the archimedean generator, it comes that
	\begin{align*}
	C(u_{1},\ldots,u_{d},s=s_{d},\Theta)
	&=\phi_{d-1,\theta_{d-1}}( \phi_{d-1,\theta_{d-1}}^{-1}\circ C \left\lbrace u_{1},\ldots u_{d-1};((s_{d-2})(d-1)),(\theta_{1},\ldots,\theta_{d-1})^{T} \right\rbrace \\
	& + \phi^{-1}_{d-1,\theta_{d-1}}(u_{d})).
	\end{align*}
	Note that the generators $ \phi_{i} $  within a HAC can come either from a single family of generators, or from different families of generators. To guarantee that C is a copula, \textcolor{blue}{McNeil \cite{B1}} provided sufficient conditions on the functions of the generator. He considers that if \[ \phi_{i} \in \Phi=\left\lbrace \phi: [0,+\infty[ \longrightarrow [0,1]  /   \phi(0)=1, \phi(\infty)=0 ;  (-1)^{j}\phi^{(j)}\geq 0; j=1, \ldots , \infty \right\rbrace; \] for $ i=1, \ldots, d-1  $ and $ \phi^{-1}_{i}\circ \phi_{i+1}   $  has a completely monotonic derivative (Sufficient nesting condition) for $ i = 1, \ldots, d-2 $, then C is a copula. However, in the case where we have different families within the same HAC model, the condition of complete monotony is not always verified.\\
	
	Generally, the structures of the HAC are arbitrary, which makes the HAC a distribution model that is both rich and varied. Moreover, the HAC can be constructed with generators belonging to different families, that is to say heterogeneous HACs or to the same family, homogeneous HACs. For a homogeneous HAC built using the same single-parameter generator function at each level, but with a different value of the $ \theta $ parameter, its entire structure is specified with $ d-1 $ parameters. The HAC approach can therefore be considered as a promising alternative whose structure and parameters must be carefully determined.
	\subsection{Tail modeling}
The extreme values theory (EVT) is mainly used in financial modeling to model the behavior of the tails of the distributions. Thus, for each residual vector $ z_ {i, t} $ obtained from the ARMA(1,1)-APARCH(1,1)-sstd model, the upper and lower tails are modeled by the generalized Pareto distribution and the central part by kernel methods as suggested by Coles (2001) used recently by \textcolor{blue}{Sahamkhadam and al \cite{B17}} and \textcolor{blue}{Sang Phu Nguyen and al \cite{C2}}. We have the following parameterization: 
	\begin{equation}
	\mathcal{G}(z_ {i, t} ) = 
	\left\{
	\begin{array}{ll}
	\frac{N_{u^{L}}}{N}\{ 1+\xi^{L}\frac{u^{L}-z_ {i, t}}{\beta^{L}}\}^{-\frac{1}{\xi^{L}}}& \mbox{if  } z_{i, t} < u^{L} \\
	\Phi(z_ {i, t})               & \mbox{if  } u^{L}< z_{i, t} < u^{R}\\
	1-\frac{N_{u^{R}}}{N}\{ 1+\xi^{R}\frac{u^{R}-z_ {i, t}}{\beta^{R}}\}^{-\frac{1}{\xi^{R}}} & \mbox{if  }  z_{i, t} > u^{R}
	\end{array}
	\right.
	\end{equation}
	where $ \xi $ is the tail index parameter, and $ \beta $ scale parameter, and $ u^{L} $ and $ u^{R} $ denote lower
	and upper thresholds, respectively.
	\section{Modeling with the ARMA-APARCH-EVT-HAC model}
	\subsection{Hierarchical dependence structure}
	In this subsection, we first use the ARMA-APARCH model to describe the behavior of each series, we will recover the residual vectors for all the series, then we specify the marginal distribution for each residual vector in equation (2) $ z_{i,t} $ with $ i=1,\ldots,d$ that is:
	\begin{equation}
	(u_{1,t},\ldots,u_{d,t})=\left( F(z_{1,t}),\ldots,F(z_{d,t}) \right). 
	\end{equation}
	where $ F (\cdot) $ represents the marginal cumulative distribution function. We then determine the underlying hierarchical structure associated with the residual vectors of the ARMA-APARCH model using the hierarchical clustering method allowing us to initiate the ARMA-APARCH-EVT-HAC modeling.\\
	
	The hierarchical regrouping and parameter estimation procedure adopted within the framework of this study is that developed by \textcolor{blue}{Jan Gorecki et al \cite{B27}}. Indeed, \textcolor{blue}{Jan Gorecki et al} have proposed an approach in which the structure and the parameters are estimated in a bottom-up manner. The technics is based on the fact that a HAC can be constructed from all of its bivariate margins and thus makes it possible to estimate the parameters of the copula only from the parameters of the bivariate marginal copula. In the context of this study, for more simplicity we use homogeneous HACs, that is to say HACs composed of ACs of the same family.\\
	
	Kendall's tau defined by the relation \ref{tau} is used to determine its structure and estimate the parameters. The generators with the highest values of Kendall's tau $ \tau $ are assigned to the lowest levels of the branches of the structure. Going higher in the tree, we assign generators with values less than $ \tau $. \\
	
	If C is a d-HAC of structure S and generators $ \Phi_ {1}, \ldots, \Phi_{d-1} $, where each parent-child pair satisfies the sufficient condition for nesting. Then, $ \tau(\phi_{i})\leq \tau(\phi_{j}) $ where $ \phi_{j} \in \mathcal{D}(\phi_{i}) $ for each $i=1, \ldots,d-1$ and $ \mathcal{D}(\phi_{i}) $ is the set of descendant nodes of $ \phi_ {i} $. \\
	
	For a parametric family and for each $ \ phi_ {k} $, the relation $ \theta \sim \tau $ is used to determine the parameters $ \theta_{k}=\tau_{\theta}^{-1}(\tau(\phi_{k})) $, for $ k=1, \ldots,d-1 $. In other words, suppose $ (u_{1},\ldots,u_{d})\sim C $ where C is a d-HAC with parametric generators $ \phi_{1}, \ldots, \phi_{d-1} $ of the same family, the structure and parameters of C can be obtained from $ (\tau_{ij}) $ using the algorithms developed by \textcolor{blue}{J. Gorecki and al \cite{B27}}. \\
	
	\subsection{VaR forecast based on ARMA-GARCH-EVT-HAC}
	\subsubsection{Computation of VaR}
	Considered to be the maximum potential loss of an investor on the value of a portfolio, the value at risk depends on three parameters: the distribution of the losses and profits of the portfolio, the level of confidence between 0 and 1 and the temporary horizon holding a portfolio of assets.\\
	
	It should be noted that in the context of this study, the VaR of a portfolio is defined by its general formulation:
	\begin{equation}
	VaR(\alpha) = F_L^{-1}(\alpha)=\inf\{ l\in \mathbb{R}:F_L(l)\geq \alpha\};
	\end{equation}
	 where $ F_L(x) = P [L \leq x] $ is the cumulative probability distribution of the loss function of random  variable L, and $ \alpha $ the conﬁdence level of the VaR.\\
	 
	 The conditional value at risk  on the probability level $ \alpha $ corresponds to the average value of the losses which exceed the VaR and it is denoted $ CVaR (X; \alpha) $. It remains more effective than VaR in terms of tail measurement. It is given by:
	 \begin{equation}
	 CVaR(X;\alpha) = \mathbb{E}[X-VaR(X;\alpha) / X > VaR(X;\alpha)].
	 \end{equation}
	 Note that the one-dimensional distribution function $ F_L $ depends on the d-dimensional distribution function of $ \log $-returns $ F_X $. Using the copula, the functions of marginal distributions $ F_{X_j} $ can be modeled separately from their dependence structure and then be coupled together to form the multivariate distribution $ F_X $.
	 \subsubsection{Estimating VaR}
	 In this subsection, we describe the modeling methodology to quantify the risks of a portfolio based on the ARMA-APARCH-EVT and HAC model. Thus, a sequential procedure composed of several steps inspired by the work of \textcolor{blue}{Maziar Sahamkhadam and al \cite{B17}} on classical copulas is developed with hierachical Archimedean copulas. In addition, the return $ r_{i, t} $, which is a time series of returns, can be modeled with the formula:
	 \[ r_{i,t}=\mu_{i}+\sigma_{i,t}\epsilon_{i,t};\]
	 where $ \mu_{i} $ is the average of the log-returns at time point t-1, $ \sigma_{i, t} $ is the square root of the associated conditional variances predicted from the APARCH model at the time point t of $ r_{i, t} $. 
	 So, the sequential procedure for modeling and predicting VaR is described as follows:
	 \newline
	 \begin{enumerate}
	 \item	Use the ARMA-APARCH model to estimate the parameters and obtain normalized residuals at time t for all series of yields:
	 	\begin{equation}
	 	\hat{\eta}= (\hat{\eta}_{1,t}, \ldots, \hat{\eta}_{d,t}),~~ \hat{\eta}_{j,t}\approx i.i.d, ~~ forall ~~  j=1,\ldots,d.
	 	\end{equation}
	 	\item Use the normalized residual vectors estimated from step 1 to specify and estimate marginal distributions $ \hat{v}_{j,t}=\hat{F}_{j}(\hat{\eta}_{j,t}) $ for $ j=1,\ldots,d$. Thus, the generalized Pareto distribution is used to parametrically model the tails of each marginal distribution and the interior by kernel methods.
	 	\item Specify the hierarchical structure of the HAC by inserting the estimated uniforms $ \hat{v}_{j,t}=\hat{F}_{j}(\hat{\eta}_{j,t}) $ for $ j=1,\ldots,d $ of step 2 and estimate the parameters by the hierarchical grouping method described in the previous section. In the case of the fully nested HAC we have:
	 	\[ C(\hat{v}_{1,t},\ldots,\hat{v}_{d,t},s=s_{d},\Theta)
	 	=C\left\lbrace \hat{v}_{1,t},\ldots, \hat{v}_{d,t}; ((s_{d-1})d),(\theta_{1},\ldots,\theta_{d-1})^{T}
	 	\right\rbrace \\ \]
	 	which is equivalent to
	 	\begin{align*}
	 	C(\hat{v}_{1,t},\ldots,\hat{v}_{d,t},s=s_{d},\Theta)
	 	&=\phi_{d-1,\theta_{d-1}}( \phi_{d-1,\theta_{d-1}}^{-1}\circ C \left\lbrace \hat{v}_{1,t},\ldots \hat{v}_{d-1,t};((s_{d-2})(d-1)),(\theta_{1},\ldots,\theta_{d-1})^{T} \right\rbrace \\
	 	& + \phi^{-1}_{d-1,\theta_{d-1}}(\hat{v}_{d,t}) ).
	 	\end{align*}
	 	\item Generate N uniform random numbers $(X_{i1}\ldots,X_{id}) $, for $ i=1,\ldots,N $ for each serie of returns $ j = 1, \ldots, d $ forming scenarios of the possible changes of the residual vector. \\
	 	
	 	The \textcolor{blue}{Monte Carlo} method generates a N number  of such scenarios, and evaluates the change in the current value of a portfolio under each scenario. If the sufficient nesting condition is fulfilled for all nodes of the nested Archimedean copula structure, then the following algorithm, based on the propositions of \textcolor{blue}{McNeil \cite{B1} and Hofert \cite{B11}}, is adopted to simulate new residues. We describe the procedure as follows:
	 	\newline
	 	\begin{itemize}
	 	\item[$ \checkmark $] Generate a variable $ V_{1} \sim F_{1}= \mathcal{LS}^{-1}\left[ \phi_{1} \right]  $ where $ \mathcal{LS}^{-1} $ is Laplace inverse transformation of the generator $ \phi_{1} $.
	 	\item[$ \checkmark $] For $ k=2, \ldots, d-1$, generate  \[ {V_{k}} \sim F_{k}= \mathcal{LS}^{-1} \left[ \phi_{k}^{(k-1)}(.;V_{k-1})\right] ~~ with~~   \phi_{k}^{(k-1)}(.;V_{k-1}) = \exp\left( -V_{k-1}\phi_{k-1}^{-1}\circ \phi_{k}(.) \right).\]
	 	\item[$ \checkmark $] Generate i.i.d $ X_{ij} \sim \exp(1)~~;~~ i=1,\ldots N $ and $ j= 1,\ldots,d $
	 	\item[$\checkmark$] For $ i=1, \ldots, N $ define the components $ \mathrm{u}_{i}=(u_{i1},\ldots, u_{id}) $, where $ u_{ij}=\phi_{j}(X_{ij}/V_{j}) $ for $ j=1,\ldots,d-1 $ and $ u_{id} = \phi_{d-1}(X_{id}/V_{d-1})$, $ u_{ij}\sim U(0,1).$
	 	\end{itemize}
	 	\item Use the uniforms simulated in step 4 $ \mathrm{u}_{i}=( u_{i1},\ldots,u_{id}) $, for $ i=1, \ldots, N $ with the determined dependence structure to estimate new normalized residuals for use in the portfolio.
	 	\begin{equation}
	 	\hat{\varepsilon}_{i}=(\hat{\epsilon}_{i1},\ldots,\hat{\epsilon}_{id})= \left( \hat{F_{1}}^{-1}(u_{i1}), \ldots, \hat{F_{d}}^{-1}(u_{id}) \right) ;~~ i=1, \ldots ,N. 
	 	\end{equation}
	 	\item Replace the estimated residual vectors $ \hat{\varepsilon}_{i} $ from step 5 in the ARMA-APARCH forecasting model and generate one-step N yield forecasts at time $ t = t_ {0} +1 $:
	 	\begin{equation}
	 	\hat{r}_{t}^{i}=(\hat{r}_{t,1}^{i}; \ldots; \hat{r}_{t,d}^{i}), for~~i=1, \ldots, N. 
	 	\end{equation}
	 	\item Use the algorithm proposed by \textcolor{blue}{Andrzej Palczewski \cite{B16}} from the PortfolioOptim package to calculate the optimal portfolio weights, \[ \hat{w}_{t}^{i} = (\hat{w}_{t,1}^{i}; \ldots ; \hat{w}_{t,d}^{i}), for~~  i=1,\ldots,N.\]
	 	\item Calculate the portfolio returns resulting from: 
	 	:\[ \hat{r}_{t,p}^{i}= \hat{w}_{t,1}^{i}\hat{r}_{t,1}^{i}+\hat{w}_{t,2}^{i}\hat{r}_{t,1}^{i}+ \ldots +  \hat{w}_{t,d}^{i}\hat{r}_{t,d}^{i}.\] 
	 	\item Use the simulated returns from step 8 and determine the value of the VaR at the corresponding time point.
	 \end{enumerate}
	 \section{Empirical study}
	 \subsection{Data}
	 The empirical data analysis performed in this section uses daily adjusted price data from US stock indices (S$\&$P500 and NASDAQ); Europeans (CAC 40 and DAX 30) provided by the “Investing.com” database and the African index (BRVM) provided by the “ww.bceao.ne” site. The raw data consists of 3,344 observations of closing values adjusted for stock market indices representative of daily transactions from January 1, 2006 to September 30, 2019, i.e. just over 13 years of observations. The greatest interest of our study relates to the returns of stock market indices whose daily returns are defined by:
	  \[ r_t =100\times \left[ (p_t-p_{t-1})/p_{t-1}\right]  \approx 100\times\left[  \ln(p_t)-\ln(p_{t-1})\right] \] 
	  with $ p_t $ the closing price of an index on day t.
	 \vspace*{1cm}
	  \begin{table}[htbp]
	  	\caption{Descriptive statistics for index returns}
	  	\label{descr}
	  	\begin{center}
	  		\begin{tabular}{|c|c|c|c|c|c|c|c|}
	  			\hline 
	  				& Max & Mean & Min & Std & Kurtosis & Skewness &  JB-test\\ 
	  			\hline
	  			S$\&$P500	& 10,9572 & 0,0245 & -9,4695 &  1,1968 & 14,3147 & -0,3723 & $18515^{\star \star \star} $ \\ 
	  			
	  			NASDAQ	& 11,1594 & 0,0367 & -9,5876 & 1.3043 & 10.2377 & -0.2943 & $7597^{\star \star \star}$ \\ 
	  			
	  			CAC40	& 10,5945 & 0,0050 & -9,4715 & 1,3737 & 9,7344 & -0,0395 & $6645^{\star\star\star} $\\ 
	  			
	  			DAX 30	& 10,7974 & 0,0235 & -7,4334 & 1,3382 & 9,2610 & -0,0438 & $ 5706^{\star\star\star} $ \\ 
	  			BRVM	& 19,8900 & 0,0068 & -19,5684 & 1,3012 & 94,4094 & -0,2228 & $1182706^{\star\star\star} $ \\ 
	  			\hline 
	  		\end{tabular} 
	  	\end{center}
	  	Notes: JB-test indicates the result of normality test \textcolor{blue}{Jarque Bera}. $ ^ {\star \star \star} $ Indicates statistical significance at the $ 1 \% $ level.
	  \end{table}
	 
	 Table \ref{descr} presents descriptive statistics of stock market index returns. Indeed, the CAC 40 index is more volatile than the other indices with a standard deviation equal to 1.3737. It should also be noted that the average returns are positive and that the NASDAQ benefited from an average of 0.0367 higher than the other averages. The skewness coefficient is a measure of the skewness of distribution of a random variable. The skewness coefficients are negative for all the series of returns, which corresponds to a spread to the left of the distribution. With regard to the kurtosis coefficient, the values of the kurtosis coefficients are much greater than 3. Thus, the tails of the empirical distributions of returns are generally thicker than those of a Gaussian (Kurtosis = 3). We then speak of the leptokurtic distribution. To be convinced of this, the normality test of \textcolor{blue} {Jarque Bera} (see \textcolor{blue}{\cite{A1}} for more details) was used. Thus, the results of the \textcolor{blue}{Jarque Bera} test reject the null hypothesis of normality for all series of returns.
	  \subsection{Marginal modeling}
	 The marginal modeling performed in this paper is based on the implementation of the ARMA-APARCH model in order to properly adjust the marginal distribution. We assume in the framework of this study that the ARMA-APARCH model has an Skew-t distribution. In addition, We performed a goodness-of-fit test based on the Akaike information criterion (AIC) to select the best model among ARMA(0,1)-APARCH(1,1)-sstd, ARMA(1,1)-APARCH(1,1)-sstd and ARMA(1,2)-APARCH(1,1)-sstd. Thus, the ARMA(0,1)-APARCH (1,1) model indicates a convergence problem for the NASDAQ return series, the ARMA(1,1)-APARCH (1,1)-sstd model presents the values ( S$\&$P500: 2.5308, NASDAQ: 2.8933; CAC40: 3.0617; DAX30: 3.0534 and BRVM: 2.2387) of the Akaike information criterion (AIC) and the ARMA(1,2)-APARCH (1,1)-sstd model presented the following AIC values (S$\&$P500: 2.5308; NASDAQ: 2.8925; CAC40: 3.0622; DAX30: 3.0537 and BRVM: 2.2387). On the basis of AIC, these last two models seem to present the same performances to model our different marginal distributions. We therefore use the ARMA(1,2)-APARCH(1,1)-sstd model to analyze our marginal series. the table \ref{ag}  summarizes the estimates of the ARMA(1,2)-APARCH(1,1)-sstd models.
	 
	  \begin{table}[htbp]
	  	\begin{center}
	  		\caption{estimation of ARMA(1,2)-APARCH(1,1) models}
	  		\label{ag}
	  		\begin{tabular}{|c|c|c|c|c|c|c|c|c|c|c|c|}
	  			\hline
	  			\multicolumn{12}{|c|}{ARMA(1,2)-APARCH(1,1)-sstd}\\
	  			\hline 
	  			Indices	& $\mu$ & ar1  & ma1  & ma2&$ \omega$ & $\alpha_1$ & $\gamma_1$  &$\beta_1$& $\delta$& skew & shape \\
	  			\hline 
	  			S$\%$P 500	&  0.11 & 0.90 & -0.95&0.03 & 0.009 & 0.11 & 0.91 & -0.85 &0.92 & 1.09 &6.4  
	  			\\  
	  			NASDAQ	&  0.13 & 0.94 & -0.97 & 0.02 & 0.01 &0.1 & 0.91 &  -0.83  & 0.98 &1.04 & 7.9 \\ 
	  			CAC 40	&  0.10 & 0.83 & -0.85 & 0.0006 & 0.006 & 0.08 & 0.93 &  -0.9 &0.79 &1.09 & 7.4 \\ 
	  			DAX 30	& 0.12 & 0.84 & -0.83 & -0.01 & 0.007 & 0.09 &  0.93 & -0.77 & 0.89 &1.05 &7.2 \\ 
	  			BRVM	& 0.04 & 0.96 & -0.94 & -0.004 &0.13 & 0.23 & 0.74 & 0.06 &  0.42 & 1.06 & 2.2 \\ 
	  			\hline 
	  			 
	  		\end{tabular} 
	  	\end{center}
	  \end{table}
	  \subsection{Impact of hierarchical modeling}
	 Risk assessment in finance is a function of several factors including the study of the strength of interaction between the variables modeling financial hazards. Thus, HACs are able to model joint distributions of asset returns by expanding the number of parameters giving rise to a more complete representation of the dependence picture. Moreover, it should be noted that the series of yields and the residues generated by these series with ARMA-APRCH-sstd models have approximately the same rank correlation matrix, that is to say the same structure of dependence. In this subsection, we build a hierarchical structure so that we can interpret each level of the hierarchy respecting the sufficient nesting condition provided by \textcolor{blue}{McNeil \cite{B1}} for each generator and its parent. This sufficient nesting condition specifically results in constraints on the parameters. Indeed, we compute the empirical version of the Kendall correlation matrix $ (\tau_{ij}^{n}) $ on the stock market data to determine the structure of the HAC. Then we use the values of ($ \tau_{ij}^{n} $) following the approach detailed in section 3.1 to estimate the parameters of the HAC. This procedure involves all the values of $ \tau_{ij} ^{n} $ bound to each node and also takes into account the value of the child parameter similar to the suggestion of \textcolor{blue}{J.G\'{o}recki and al \cite{B27}}.\\
	 
	  \begin{table}[htbp]
	  	\begin{center}
	  		\caption{Kendall correlation matrix}
	  		\label{cor}
	  		\begin{tabular}{|c|ccccc|}
	  			\hline 
	  			&	S$\%$P 500  & NASDAQ & CAC 40 & DAX 30 & BRVM  \\ 
	  			\hline 
	  			S$\%$P 500	&1.0000  & 0.6978 & -0.0012 & 0.0350  & 0.0014 \\ 
	  			
	  			NASDAQ	& 0.6978 & 1.0000 & 0.0048 & 0.0369 & -0.0010 \\ 
	  			
	  			CAC 40	& -0.0012 &  0.0048 & 1.0000 & 0.0144 & 0.0032 \\ 
	  			
	  			DAX 30	& 0.0350 &  0.0369 & 0.0144 & 1.0000 & -0.0027 \\ 
	  			
	  			BRVM	& 0.0014 & -0.0010 & 0.0032 & -0.0027 & 1.0000  \\ 
	  			\hline 
	  		\end{tabular} 
	  	\end{center}
	  \end{table}
	  
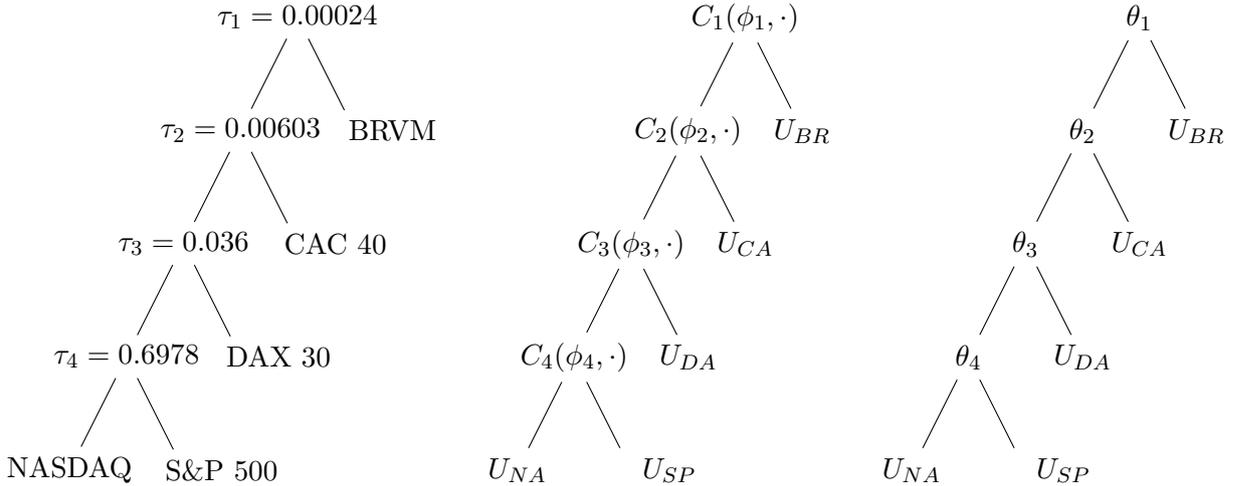
\begin{figure}
	  	\begin{center}
	  		\begin{tabular}{ccc}
	  			\begin{tikzpicture}
	  			\node{$\tau_{1}=0.00024$}
	  			child {node {$\tau_{2}=0.00603$}child {node {$\tau_{3}=0.036$}
	  					child {node {$\tau_{4}= 0.6978$} child {node {NASDAQ}}
	  						child {node {\hspace*{1cm}S$\&$P 500}}}
	  					child {node {\hspace*{1cm}DAX 30}}}
	  				child {node {\hspace*{1cm}CAC 40}} }
	  			child {node {\hspace*{1cm}BRVM}
	  			};
	  			\end{tikzpicture}	&  \begin{tikzpicture}
	  			\node{$C_{1}(\phi_{1},\cdot)$}
	  			child {node {$C_{2}(\phi_{2},\cdot)$}child {node {$C_{3}(\phi_{3},\cdot)$}
	  					child {node {$C_{4}(\phi_{4},\cdot)$} child {node {$U_{NA}$}}
	  						child {node {\hspace*{1cm}$U_{SP}$}}}
	  					child {node {$U_{DA}$}}}
	  				child {node {$U_{CA}$}} }
	  			child {node {$U_{BR}$}
	  			};
	  			\end{tikzpicture}  &   \begin{tikzpicture}
	  			\node{$\theta_{1}$}
	  			child {node {$\theta_{2}$}child {node {$\theta_{3}$}
	  					child {node {$\theta_{4}$} child {node {$U_{NA}$}}
	  						child {node {\hspace*{1cm}$U_{SP}$}}}
	  					child {node {$U_{DA}$}}}
	  				child {node {$U_{CA}$}} }
	  			child {node {$U_{BR}$}
	  			};
	  			\end{tikzpicture}  \\ 
	  		\end{tabular} 
	  	\end{center}
	  	\caption{Hierarchical structure of stock market indices. The tree on the right contains the abbreviations NASDAQ (NA);  S$\&$P 500 (SP); DAX 30 (DA); CAC 40 (CA) et BRVM (BR).}
	  	\label{struc}
	  \end{figure}

	 It emerges from the analysis of the Kendall correlation matrix (table \ref{cor}) that the US S$\&$P500 and NASDAQ indices are strongly correlated, given that they are tied to the same economic environment and some stocks are present in both these indices, for example, Apple Inc. (AAPL) which is listed on the S$\&$P500 and the Nasdaq. We therefore assign these two indices to the lowest level 4 of the structure. For these two indices, a significant geographic effect is observed. Kendall's correlation between the DAX 30 and the American indices is more significant than their correlation with the CAC 40 or BRVM. This is why we have assigned DAX 30 to level 3 of the structure. CAC 40 and BRVM are respectively grouped together on levels 2 and 1 of the structure. The figure \ref{struc} visually illustrates this structure. the correlation of kendall $ \tau_{1} $ denotes the average correlation between the BRVM index and the other indices, this positive correlation being weak this reflects the fact that BRVM is negatively correlated with NASDAQ and DAX 30; $ \tau_{2} $ denotes the average correlation between the CAC 40 index and the indices (DAX 30, NASDAQ and S$\& $P 500) with a negative correlation with S$\&$P 500 and $ \tau_{3} $ denotes the average kendall correlation between the DAX 30 index and the indices (NASDAQ and S$\&$P 500). The structure obtained following this regrouping is given by the relation:
	  \begin{equation}
	  C(U_{SP},U_{NA},U_{DA},U_{CA},U_{BR})
	  =C_{1}\left\lbrace C_{2}(C_{3}(C_{4}(U_{NA},U_{SP}),U_{DA}),U_{CA}) ,U_{BR}\right\rbrace \\
	  \end{equation}
	  by using the generator,
	  \begin{equation}
	   C(U_{SP},U_{NA},U_{DA},U_{CA},U_{BR})=\phi_{1}\left\lbrace \phi_{1}^{-1}\circ C_{2}(U_{SP},U_{NA},U_{DA},U_{CA})+\phi_{1}^{-1}(U_{BR}) \right\rbrace.
	  \end{equation}
	  We use the means of the pairwise versions of Kendall's tau and the relationship $ \theta\sim \tau $ to estimate the different parameters of the HAC. Note that this HAC specification will be used throughout the simulation study of the new residues. The most popular copulas of the Archimedean family are used for this estimate namely Gumbel's and Clayton's copula which are asymmetric functions and reflect the dependence of the upper tail and lower tail respectively, while the Frank one is symmetric but without accounting for tail dependence. Joe's copula is similar to Clayton's copula.
	  \begin{table}[htbp]
	  	\caption{HAC parameter estimates}
	  	\begin{center}
	  		\begin{tabular}{|ccccc|}
	  			\hline 
	  			Structures	& $\theta_{1}$ & $ \theta_{2} $ & $ \theta_{3} $ & $ \theta_{4} $ \\ 
	  			\hline 
	  			HAC-Frank	&  0.0022 &  0.0543 & 0.3243 &  11.3157 \\ 
	  			
	  			HAC-Joe	& 1.0004 & 1.0105 & 1.0647 & 5.4178 \\ 
	  			
	  			HAC-Gumbel	&  1.0002 & 1.0060 & 1.0373 & 3.3100\\ 
	  			
	  			HAC-Clayton	&  0.00049 &  0.0121 &   0.0747 &   4.6201
	  			\\ 
	  			\hline 
	  		\end{tabular} 
	  	\end{center}
	  \end{table}
	  \subsection{Results}
	 To prepare our approach to the hierarchical Archimedean copula in quantifying the risks of a financial portfolio, as we described in section 3.2.2. We must first specify and estimate the marginal distributions that better correspond to the residuals obtained from the ARMA-APARCH-sstd model. Indeed, the residual series have heavier tails than the normal one. To take this fact into account, we use the semi-parametric approach. So, we construct a piecewise marginal distribution by parametrically modeling the tails of each residual vector by the generalized Pareto distribution (GPD) and the interior by kernel methods. In this way, we determine the appropriate upper and lower thresholds such that 10$\% $ of residuals are reserved for each tail, then we estimate the tail index and scale parameters of the GPD. We use the uniforms estimated from the marginal residual distributions to estimate the structure of the HAC. Next, we simulate 10,000 uniforms that we insert into the inverse function of the marginal distribution to obtain new residuals. Then, we incorporate these simulated residual vectors into the ARMA-APARCH model to jointly generate series of yields. These simulated returns are used to examine the risks of each portfolio and assess their performance.\\
	  
	  The simulated returns are used through the optimization algorithm proposed by \textcolor{blue}{Andrzej Palczewski \cite{B16}} to determine the optimal weights for each portfolio and estimate the risk measures. This study aims to compare the performance of ARMA(1,2)-APARCH(1,1)-EVT prediction models based on Archimedean copulas (ACs) and hierarchical Archimedean copulas (HACs) in terms of measurements risk. However, it should be noted that the choice of model depends on the expectations of investors. Thus, value-at-risk (VaR), conditional value-at-risk (CVaR) and average run time (ART) are evaluated in this document. In all calculations, we consider the confidence level $ \alpha = 95\% $. \\
	  
	 The \ref{VaR} table summarizes the results of the risk measurements and the average run times of the different models that generate our portfolios. In terms of CVaR, it is clear that the portfolios generated using the ARMA-APARCH-EVT-HAC model reported lower values. However, the observed VaRs indicate various results. Indeed, the HAC-Clayton and the HAC-Frank generate portfolios with lower VaRs compared to their ACs, which is not the case for the copulas of Joe and Gumbel. These results promise better prospects for risk averse investors. In general, the results of this study suggest that risk averse investors should use the ARMA-APARCH-EVT-HAC model rather than an ARMA-APARCH-EVT-AC model to manage the risks of their portfolios. \\
	  
	  The last column of the \ref{VaR} table contains the average run times (ART) of the 10 different realizations aimed at simulating our different portfolios. It should be noted that our objective is not to analyze the conciseness of the average run time of the samples as it mainly depends on the speed of the processor used but rather we try to compare these average run times of the different models according to the copula class used. Thus, the ART corresponds within the framework of our study to the average time that the machine will take to simulate new residues based on copulas and then incorporate them into the ARMA-APARCH model and simulate yield samples. the average time is calculated after 10 trials. It follows from the analysis made of the results obtained that the execution times of the samples generated by the EVT-HACs models are slightly higher than the execution times of the samples generated by the EVT-ACs model. This is explained by the fact that the HAC model is an assembly of several ACs.\\
	  
	  Finally, a global analysis of the results of table \ref{VaR} allows us to affirm that the ARMA-APARCH-EVT-HAC models produce less risky portfolios with naturally a slightly higher calculation load than that of the ARMA-APARCH-EVT-AC models. \\
	  
	  \begin{table}[htbp]
	  	\begin{center}
	  		\caption{VaR and CVaR and  estimate from copulas (in \%). Average run time ART(in sec.)}
	  		\label{VaR}
	  		\begin{tabular}{|l|c|c|c|c|c|}
	  			\hline 
	  			portfolios&Forecast model& Copula	& VaR & CVaR   & ART  \\ 
	  			\hline
	  			(P1)&  &Clayton	& 1.387 &  2.677 &   0.922
	  			\\ 
	  			
	  			(P2)&  & HAC-Clayton	& 1.334 &   2.585 &  1.312 \\ 
	  			
	  			(P3)& ARMA(1,2)- & Frank	& 1.421 & 2.738 &  0.971 \\ 
	  			
	  			(P4)& APARCH(1,1)- & HAC-Frank	& 1.281 & 2.416 &   1.016  \\ 
	  			
	  			(P5)& EVT & Joe & 1.331  &  2.993  &1.042  \\ 
	  			
	  			(P6)& & HAC-Joe	& 1.336 & 2.812 & 0.963 \\ 
	  			
	  			(P7)& & Gumbel	& 1.311  &   2.951  &  1.015
	  			\\ 
	  			
	  			(P8)& & HAC-Gumbel	&1.337& 2.590 &   1.076  \\ 
	  			\hline
	  		\end{tabular} 
	  	\end{center}
	  \end{table}
	  
	   \subsection{Back-testing for VaR}
	   VaR estimates produced by our various portfolios must be submitted to a validation test to examine their exactitude. In this paper, we apply the back-testing procedure widely used in quantitative risk management. Indeed, the validation of the VaR modelization requires that the percentage of exceedances be very close to the predefined VaR level and that the null hypothesis of the independence of the exceedances not be rejected. We can test these properties by running the unconditional coverage test of \textcolor{blue}{Kupiec} see \textcolor{blue}{\cite{B8}} or \textcolor{blue}{\cite{B9}}, which compares the exceedance rate at the predefined level of VaR and the independence test of \textcolor{blue}{Christoffersen and Pelletier see \cite{A4}}, which checks the independence of the exceedances. The table \ref{bac} presents the results of the unconditional coverage (UC) and conditional coverage (CC) tests with a confidence level $ \alpha = 95\% $. The p-values obtained with our experimental data indicate that all copula models have particularly shown good performances and satisfactory precision, which is particularly important in the quantitative management of risks in finance. However, based on the results of two back-testing procedures, it is difficult to say that the ARMA-APARCH-EVT model based on hierarchical Archimedean copulas gives better estimates. Nonetheless, the HAC model for Clayton outperforms all other models examined. In addition, the analysis of the results on the proximity of the level of VaR exceedance to the expected level of exceedance (i.e. the proximity of $ 1- \alpha = 5\% $) shows that the HAC model with the Clayton and Frank generators, and Joe's copula provide an overshoot level very close to the preset level. We can therefore conclude that these models produce acceptable and more precise measures of risk compared to other models.
	  \begin{table}[htbp]
	  	\begin{center}
	  		\caption{back-testing results}
	  		\label{bac}
	  		\begin{tabular}{|l|c|c|c|c|c|}
	  			\hline 
	  			Portfolios& Forecast Model& Copula	& UC p-value & CC p-value & exceedances $95\% $ \\ 
	  			\hline
	  			(P1)&  &Clayton	& 0.204 & 0.428  &  5.9\% 	\\ 
	  			
	  			(P2)&  & HAC-Clayton&   0.884	& 0.917 &  4.9\%   \\ 
	  			
	  			(P3)& ARMA- & Frank	&  0.557  &  0.702 &   4.6\%   \\ 
	  			
	  			(P4)& APARCH- & HAC-Frank	& 0.77 & 0.54 &  4.8\% \\ 
	  			
	  			(P5)& EVT & Joe &0.885 & 0.91  &5.1\%  \\ 
	  			
	  			(P6)&  & HAC-Joe	& 0.299  & 0.579 &  4.3\% \\ 
	  			
	  			(P7)& & Gumbel	&  0.566  &  0.709  &    5.4\% 	\\ 
	  			
	  			(P8)& & HAC-Gumbel	&  0.097  & 0.235 &  3.9\%      \\ 
	  			\hline
	  		\end{tabular} 
	  	\end{center}
	  \end{table}
	  \section{Conclusion and Discussion}
	 In this work, we have proposed a new approach of applicability of flexible models of HAC in the methodology of existing modeling. A careful study of the hierarchical dependence structure between international stock market indices was discussed. The empirical results reveal the need to use the HAC to precisely analyze the interactions between financial series. \\
	  
	  The ARMA-APARCH volatility analysis model was combined with extreme value theory (EVT) to construct a marginal distribution by parametrically modeling the tails of each residual vector by the generalized Pareto distribution (GPD). Thus, the HAC model is used to simulate new residues to be incorporated into the ARMA-APARCH model and jointly simulate the series of yields. We used these simulated returns to calculate the optimal weights for each portfolio and estimate the risk measures. The traditional back-testing approach is used to select the best model. In addition, the performance of portfolios generated by HACs and ACs was analyzed and discussed. The portfolio risk assessment results based on value at risk (VaR) and conditional value at risk (CVaR) show that the ARMA-APARCH-EVT-HAC model is more efficient due to its ability to generate less risky portfolios.However, based on the overall test results of \textcolor{blue}{Kupiec} and \textcolor{blue}{Christoffersen and Pelletier}, the copula models provided mainly better risk estimates. But in terms of average execution time, the HAC-based model has a slightly higher computational load than the AC-based model.\\
	  
	  Finally, we can say that with an appropriate marginal distribution and an adequate specification of the HAC model, the proposed modeling process can contribute to significantly reduce the risk of the portfolio.\\
	  
	   Moreover, to the contribution of this study, further investigations could be conducted with nested copula time varying and other forecasting models volatility.

	  \bibliographystyle{plain}

\end{document}